\documentclass[twocolumn,reprint,amsmath]{revtex4-1}
\usepackage{amsmath}
\usepackage{txfonts}
\usepackage{hyperref}
\usepackage{graphicx}
\usepackage{float}
\usepackage{color}
\usepackage[normalem]{ulem}
\usepackage{todonotes}
\usepackage[utf8]{inputenc}

\DeclareMathOperator{\LT}{\mathcal{L}}
\DeclareMathOperator{\Surva}{S_{\alpha}}
\DeclareMathOperator{\ga}{g_{\alpha}}
\DeclareMathOperator{\lga}{\tilde{g}_{\alpha}}

\DeclareMathOperator{\lrho}{\tilde{\rho}_{0}}
\DeclareMathOperator{\Surv}{S}
\DeclareMathOperator{\lSurv}{\tilde{S}}
\DeclareMathOperator{\lSurva}{\tilde{S}_{\alpha}}

\DeclareMathOperator{\Tm}{\overline{T}}
\DeclareMathOperator{\Tma}{\overline{T}_{\alpha}}
\DeclareMathOperator{\Ta}{T_{\alpha}}
\DeclareMathOperator{\To}{T_{0}}

\newcommand{\dd}{\mathrm{d}}
\newcommand{\scale}{\tau_0}
\newcommand{\tdiff}{\tau_{\mathrm{diff}}}
\newcommand{\lk}[1]{\textcolor{black}{#1}}

\begin{document}

\title{Robust random search with scale-free stochastic resetting}
\author{\L{}ukasz Ku\'smierz}
\affiliation{Laboratory for Neural Computation and Adaptation, RIKEN Center for Brain Science, 2-1 Hirosawa, Wako, Saitama 351-0198, Japan}
\author{Taro Toyoizumi}
\affiliation{Laboratory for Neural Computation and Adaptation, RIKEN Center for Brain Science, 2-1 Hirosawa, Wako, Saitama 351-0198, Japan}

\begin{abstract}
	A new model of search based on stochastic resetting is introduced, 
    wherein rate of resets depends explicitly on time elapsed since the beginning of the process. 
    It is shown that rate inversely proportional to time leads to paradoxical diffusion 
    which mixes self-similarity and linear growth of the mean square displacement with 
    non-locality and non-Gaussian propagator.
	It is argued that such resetting protocol offers a general and efficient search-boosting method 
	that does not need to be optimized with respect to the scale of the underlying search problem 
    (e.g., distance to the goal) and is not very sensitive to other search parameters. 
	Both subdiffusive and superdiffusive regimes of the mean squared displacement 
	scaling are demonstrated with more general rate functions. 
\end{abstract}
    \maketitle
	\section{Introduction}
    Any search is intrinsically governed by randomness: 
    the need to perform it underscores the ignorance of the searching agent 
    and its inability to predict the outcome. 
	In particular, the agent may not know in advance how much time, 
	on average, it will take to find a target and 
	how much variability in search time is expected due 
	the randomness. 
	How can the efficiency of a search strategy be boosted 
    without acquiring such information beforehand? 
    It may happen that the longer the search lasts, 
	the more unfavorable random instance the agent is experiencing, 
	and thus the longer the expected remaining time to find the target. 
	It is then true that restarting the process anew may be beneficial, 
	as it lowers the expected completion time. 
	It has been shown that this is indeed the case if 
	the coefficient of variation of completion times 
	is larger than $1$ \cite{Reuveni2016}. 
	This conditions holds if the completion times of a search process 
	are drawn from the hyperexponential or power-law distributions. 
	Such situation arises, for instance, in the one-dimensional, 
	unbiased Brownian motion with a single target. 
	The possibility of arbitrarily long excursions far from the target 
	manifests itself in large fluctuations and the associated 
	power-law distribution of the first passage times.
	
	In the engineering community such mechanism was studied 
	in the context of computer software systems 
    \cite{huang1995software,dohi2001estimating,van2006analysis,okamura2016analysis} 
	and non-convex optimization methods 
	\cite{fukunaga1998restart,jansen2002analysis,luersen2004globalized,loshchilov2016sgdr}. 
	More recently, Evans and Majumdar introduced a model 
    with time-homogeneous stochastic resetting 
    (i.e., with exponentially distributed waiting times between the resets)
    in tandem with diffusion 
	\cite{majumdar2011resetting}. 
	They showed that one-dimensional \cite{majumdar2011resetting} and 
	multidimensional \cite{evans2014diffusion} diffusion with resets 
	exhibit finite mean first passage times (MFPTs) 
	which can be optimized with respect to resetting rate $r$. 
	Moreover, resets lead to non-equilibrium steady-states 
	with a combination of local and non-local currents 
	\cite{majumdar2011resetting,mendez2016characterization,eule2016non}. 
	The simplicity and nontrivial behavior of these systems 
	has sparked the interest of the physics community, 
	with a considerable amount of research focusing on 
	modifying the model to include nontrivial boundary conditions 
	\cite{christou2015diffusion,chatterjee2018diffusion,pal2018first}, 
	external potentials \cite{pal2015diffusion},
	anomalous transport 
	\cite{kusmierz2014first,kusmierz2015optimal,mendez2016characterization,shkilev2017continuous,Montero2017,bodrova2018scaled,bodrova2018non,kusmierz2019subdiffusive}, 
	drift \cite{Montero2017,ray2018p}, 
	delays following the resets \cite{Reuveni2016, evans2018effects, maso2019stochastic, pal2019home},
	and
	non-exponential distributions of waiting times between resets 
	\cite{nagar2016diffusion,eule2016non,pal2016diffusion}.
	Resets were also studied in the context of 
	stochastic energetics \cite{fuchs2016stochastic}, 
	enzymatic reactions \cite{rotbart2015michaelis}, 
	fluctuating interfaces \cite{gupta2014fluctuating}, 
    and power-law distributions in non-equilibrium systems  
	\cite{manrubia1999stochastic}.

	Non-exponential waiting times between the resets can be generated 
	by time-dependent resetting rates 
	\cite{nagar2016diffusion,pal2016diffusion}.
	In models introduced in these previous works, 
	rate depends on the time elapsed since the last reset. 
	Here we introduce a new model with resetting rate $r(t)$ that
	depends on the time elapsed since the start of the process. 
    Such non-stationary resetting protocol, 
    while preserving the Markovian character of the dynamics,   
    introduces time-inhomogeneity.   
	As we will show, this dramatically affects the behavior of the process. 
	If $r(t)$ decays with time, the process does not converge to any steady 
	state, with the mean square displacement (MSD) growing indefinitely. 
	Moreover, $r(t)$ that is inversely proportional to time provides an efficient 
	and robust mechanism to boost the time efficiency of a search. 
	It has previously been shown that deterministic resetting 
	is the optimal way of minimizing search time via resets \cite{pal2016first}. 
	However, similarly to stochastic resetting with a constant rate, 
	such resetting has to be optimized assuming full knowledge 
	of details of a given search problem. 
	In contrast, the resetting mechanism we propose does not have to be 
	adapted to the time scale of the search problem. 

	\section{Diffusion with time-dependent resetting}
    We start from one-dimensional diffusion as the underlying search process, 
    which will help build the intuition. 
    Later we present a more general framework. 
    Since the process of diffusion with memoryless resetting is Markovian, 
    it is fully characterized by the initial position  
    distribution and the propagator $\rho(x,t|x_i,t_i)$,  
    which denotes the probability density function of $x$ at time $t$ 
    of a particle that at time $t_i<t$ was at $x_i$. 
	The particle diffuses till a resetting event that
	brings it instantaneously back to the resetting position $x_0$ 
	(equal to the position of the particle at time $t=0$)
    \footnote{
    We are assuming that the initial position corresponds to the resetting position.
    However, due to the singular behavior of the function $r(t)$ at $t=0$, 
    the choice of the initial position is inconsequential:
    If it does not match the resetting position,
    the probability mass is transferred to the resetting position
    in an infinitesimal time.
    }.
    The particle then continues to diffuse until the next reset.
	The corresponding propagator solves the following partial differential equation
    \begin{equation}
        \partial_t \rho(x,t|x_i,t_i) = 
        D \partial^2_{xx} \rho(x,t|x_i,t_i) - r(t) \rho(x,t|x_i,t_i) + r(t) \delta(x-x_0)
        \label{eq:general-diff-free}
    \end{equation}
    where $\partial_z$ denotes the partial derivative with respect to $z$ and 
    $D$ is the diffusion coefficient.
    Such process has been analyzed extensively in the case of a constant rate $r(t)=r$, 
    with the corresponding resets described by the homogeneous Poisson point process.
    It has been shown \cite{majumdar2011resetting} that this process, 
	in the absence of targets, attains a non-equilibrium steady state.  
	If a single target, represented by an absorbing boundary, is placed at the origin, 
	there exists an optimal rate $r^*\propto x_0^{-2}$ at which the MFPT is minimized. 
	
	\subsection{Scale-free resetting}
    In this work we mainly focus our attention on the special case of 
    \begin{equation}
        r(t)=\frac{\alpha}{t}.
        \label{eq:r-propto-inv-t}
    \end{equation} 
    The only parameter $\alpha$ is dimensionless---this form of resetting does not 
    introduce any time scale. 
    For this reason, we refer to (\ref{eq:r-propto-inv-t}) as a \textit{scale-free resetting}.
    Resets described by (\ref{eq:r-propto-inv-t}) are generated by 
    the inhomogeneous Poisson point process 
    with an average intensity (expected number of events) within a 
    time period $(t_0,t_1]$ given by
    \begin{equation}
        N(t_0,t_1) = \alpha \ln\frac{t_1}{t_0}.
    \end{equation}
    Note that our choice of $r(t)$ features a singularity at $t=0$, 
	which translates into diverging intensity with $t_0\to 0$. 
	Any practical applications of scale-free resetting have to 
	introduce a short-term cut-off to avoid resetting with infinite frequency.  
	However, all our theoretical predictions without an explicit account of the 
	cut-off are still valid at times much longer than the cut-off. 

\begin{figure}
		\includegraphics[width=0.99\linewidth]{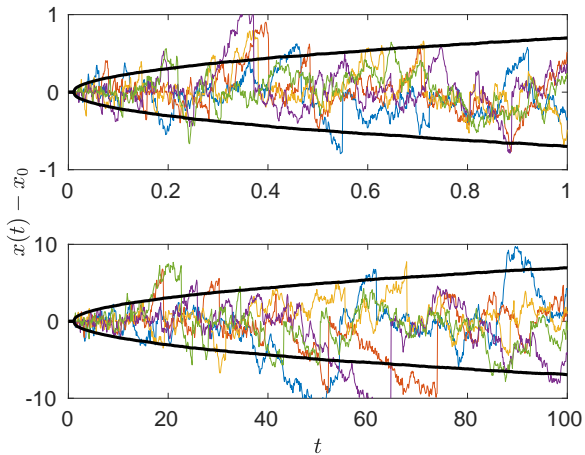}
		\caption{
			Sample trajectories (colored, thin lines) and percentile lines 
			(thick, black lines) 
			of diffusion with scale-free resetting with $\alpha=10$.
			A comparison of short (top) and long (bottom) time scales illustrates 
			the self-similarity of the process.	
			The presented percentile lines $x_q(t)$ are such that 
			together only $10\%$ of the sample trajectories have $x(t)>x_q(t)$ 
			(upper branch) or $x(t)<x_q(t)$ (lower branch). 
		}
		\label{fig:quantiles}
	\end{figure}

	As evident from (\ref{eq:general-diff-free}), 
    the process (\ref{eq:general-diff-free}) with (\ref{eq:r-propto-inv-t}) 
    preserves the self-similarity of the pure Brownian motion, 
    i.e., the re-scaling $x\to c x$ and $t\to c^2 t$ does not change any observables 
	(see Fig.~\ref{fig:quantiles}). 
	We can therefore expect that the mean square displacement 
    (MSD) scales linearly with time. 
    Indeed, straightforward calculations 
	(multiply (\ref{eq:general-diff-free}) by $(x-x_0)^2$, 
	integrate over $x$ and solve the resulting ordinary differential equation)
	lead to
	\begin{equation}
        \left\langle (x(t) - x_0)^2\right\rangle 
        =
        \int\limits_{-\infty}^{\infty}dx x^2 \rho(x,t|x_0,0)
        = 
        \frac{2 D t}{1+\alpha}.
    \end{equation}
    Although the self-similarity and associated linear time-dependence of the MSD 
    bear a strong resemblance to free diffusion, 
    resets vastly modify other statistics. 
	For instance, the Fourier transform of the propagator reads
    \begin{equation}
        \rho(k,t|x_0,0) = 
        \alpha e^{i k x_0} 
        t^{-\alpha} e^{-D k^2 t} 
        \int\limits_0^t d\tau \tau^{\alpha-1} e^{D k^2 \tau}. 
    \end{equation}
    This propagator is clearly non-Gaussian and has a cusp at ${x=x_0}$. 
    It takes a particularly simple form for $\alpha=1$ 
    \begin{equation}
        \rho(k,t|x_0,0) = 
        e^{i k x_0} \frac{1 - e^{-D k^2 t}}{D k^2 t}.
    \end{equation}
	\lk{
	Similar non-Gaussian displacement distributions with a cusp, 
	accompanied by a linear time dependence of the MSD have been observed in 
	several systems of diffusing colloidal particles 
	\cite{wang2009anomalous,wang2012brownian}. 
	Such systems have recently attracted considerable attention among theoreticians, 
	who since then have developed multiple models with fluctuating diffusivity 
	\cite{hapca2008anomalous, wang2009anomalous, chubynsky2014diffusing,chechkin2017brownian,sposini2018random}.
	Scale-free resetting provides an alternative mechanism behind such behavior. 
	}
	
	\subsection{Generalizations of the resetting rate function}
	\label{sec:generalizations1}
    In addition to the presented applications to search problems, 
    a generalization of the proposed resetting mechanism can be applied 
    to model anomalous transport phenomena. 
    The MSD of (\ref{eq:general-diff-free}) 
	with $r(t)=\alpha/t^{\mu}$ is given by the formula
    \begin{equation}
        \left\langle \left( x(t) - x_0\right)^2 \right\rangle = 
        2 D \exp\left(-\frac{\alpha}{1-\mu}t^{1-\mu}\right)
        \int\limits_0^t d\tau \exp\left(\frac{\alpha}{1-\mu}\tau^{1-\mu}\right),
        \label{eq:msd-mu}
    \end{equation}
    which for $\mu<1$ exhibits a smooth transition between 
    diffusive behavior with the MSD ${\approx~2 D t}$ 
    for ${t\ll\tau_{\alpha}}$ 
    and subdiffusive behavior with the MSD ${\approx 2D t^{\mu}/\alpha}$ 
    for $t\gg\tau_{\alpha}$, 
    where ${\tau_{\alpha} = \alpha^{-1/(1-\mu)}}$ is the time scale 
    introduced by the power-law resetting. 
	In the opposite case of $\mu>1$ the long-term behavior is diffusive, 
    whereas at short times an apparent superdiffusivity 
    with the MSD $\approx 2D \tau^{\mu}/\alpha$ is observed. 
	By assuming $r(t) \propto (\log t)^{-1}$ one can also model 
	an ultra-slow diffusion with the MSD $\propto \log t$,
	a behavior previously uncovered in the strongly non-Markovian random walks with
	preferential relocations to places visited in the past \cite{boyer2014random}. 
	
	One could involve heavy-tails jump distributions leading to a competition between 
	superdiffusivity of the L\'evy flights and subdiffusive tendency from resets, 
	similar to the competition observed in the continuous-time random walk scheme 
	\cite{montroll1965random,metzler2000random,magdziarz2007competition}. 
    Many different variants of a combination of anomalous transport 
	with constant-rate stochastic resetting were previously studied 
	\cite{kusmierz2014first,kusmierz2015optimal,mendez2016characterization,shkilev2017continuous,Montero2017,kusmierz2019subdiffusive} 
	and were shown to exhibit non-trivial features, 
	including first and second order transitions of the optimal search 
	\cite{kusmierz2014first,kusmierz2015optimal,campos2015phase,Montero2017}
	and a non-monotonic behavior of the MSD \cite{kusmierz2019subdiffusive}. 
	For this reason, we expect that a combination of scale-free resets with 
	independent constant-rate resets may lead to thought-provoking phenomena.
	
	\section{Completion times}
    In the previous section we have shown how time-dependent 
	resetting affects the time evolution and shape of the displacement distribution. 
	Another striking consequence of scale-free resetting 
    can be observed in the statistics of the first passage times. 
    As we show in the following, in contrast to free diffusion 
	on infinite line, diffusion with scale-free resetting can find 
	a target in a finite MFPT.
    Moreover, due to the self-similarity, we can expect that the MFPT is 
	proportional to $x_0^2$, where $x_0$ denotes the initial 
	position with respect to the target. 
	
	This observation prompted us to explore more general scenarios of 
	search under scale-free resetting beyond diffusion on intinite line. 
	Since the following calculations apply to general search times that may not be 
	generated as the first passage times of some stochastic process, 
	we hereafter use the term \textit{completion times}.
	We show that scale-free resetting is \textit{robust}, 
	i.e. the optimal form of the scale-free resetting 
	(as prescribed by the parameter $\alpha$) does not depend on the time scale. 
	Thus, the protocol can be optimized in a parsimonious manner, 
	without knowing how much time on average the underlying search takes.
	
	\subsection{General analysis}
    Instead of assuming that the underlying search process is described by a simple, 
    one-dimensional diffusion, we first consider any \lk{random} search process\lk{. 
    Let a non-negative random variable $\To$ 
    be the completion times of the original process without resetting. 
    Similarly, let $\Ta$ denote the completion times of the same process 
    modified by introducing scale-free resets of the form (\ref{eq:r-propto-inv-t}).
    Our goal is to calculate statistics of $\Ta$ given known statistics 
    of $\To$. 
    In particular, we will be interested in the mean completion time (MCT), 
	denoted as $\Tma$. 
    }
	For any absolute-time-dependent rate function $r(t)$,  
	one can link the survival probability of the model with resets $\Surva(t)$ 
	with the survival probability of the underlying reset-free process $\Surv_0(t)$
	by means of the following renewal equation 
	\footnote{
	Note that multiple different renewal approaches to processes with resetting 
	have been proposed 
	\cite{gupta2014fluctuating,Reuveni2016,pal2016first,chechkin2018random}. 
	Our approach is closely related to a discrete-time 
	counterpart applied in \cite{kusmierz2014first}. 
	}
    \begin{equation}
	\Surva(t) = 
	\Surv_0(t) \Psi(0,t) 
	+ 
	\int\limits_0^t d\tau 
	r(\tau) \Surva(\tau) \Surv_0(t-\tau) 
	\Psi(\tau,t),
	\label{eq:renewal-general}
	\end{equation}
	where 
	\begin{equation}
	\Psi\left(t_0,t_1\right) = 
	\exp\left(-\int\limits_{t_0}^{t_1} 
	d\tau r(\tau)
	\right) 
	\label{eq:psi_general}
	\end{equation}
	is the probability of no resets in the interval of time $(t_0,t_1]$. 
	Eq.~(\ref{eq:renewal-general}) 
	splits the survival probability into two cases:
	either there are no resets until time $t$, which happens 
	with probability $\Psi(0,t)$, or there are resets in this time period 
	with the last reset at time $\tau$. 
	The integral appears here because $\tau$ can be anywhere between $0$ and $t$. 
	In the case of scale-free resetting (\ref{eq:r-propto-inv-t}), 
	the \lk{no-reset} probability is given by
	\begin{equation}
	\Psi_{\alpha}\left(t_0,t_1\right) =
	\left(\frac{t_0}{t_1}\right)^{\alpha}.
	\label{eq:psi_alpha}
	\end{equation}
	Combining the general expression (\ref{eq:renewal-general}) 
	with
	(\ref{eq:psi_alpha}) 
	and
	(\ref{eq:r-propto-inv-t}) 
	we arrive at the equation 
	\begin{equation}
	\Surva(t) = 
	\frac{\alpha}{t^{\alpha}} 
	\int\limits_0^t d\tau 
	\tau^{\alpha-1}
	\Surva(\tau)
	\Surv_0(t-\tau).
	\label{eq:renewal-scale-free}
	\end{equation}
	In order to solve Eq.~(\ref{eq:renewal-scale-free}) 
	we introduce an auxillary function
	\begin{equation}
		\ga(t) \equiv 
		t^{\alpha-1} \Surva(t),
		\label{eq:g-def}
	\end{equation}
	which solves a simpler integral equation
	\begin{equation}
		\ga(t) = 
		\frac{\alpha}{t}
		\int\limits_0^t d\tau 
		\ga(\tau) \Surv_0(t-\tau)
		\label{eq:renewal-g}.	
	\end{equation}
	By denoting ${\LT\{f(t)\} \equiv \tilde{f}(s)}$ 
	as the Laplace transform of $f(t)$, 
	the corresponding Laplace-transformed equation reads
	\begin{equation}
		\partial_s \lga(s) = 
		- \alpha \lSurv_0(s) \lga(s).
		\label{eq:diff-eq-g}
	\end{equation}
	\lk{In the following we will assume that 
	the probability of finding the target in no-time is equal to zero. }
	In this case the corresponding survival probability can be written in the form 
	\lk{\cite{redner2001guide}}
	\begin{equation}
		\lSurv_0(s) = 
		\frac{1 - \lrho(s)}{s},
		\label{eq:surv-from-dens}
	\end{equation}
	where $\rho_0(t)$ is the probability density function 
	of the completion time of the process without resetting, 
	and ${\lim_{s\to\infty}\lrho(s) = 0}$.
	We can easily solve the differential equation 
	(\ref{eq:diff-eq-g}) leading to 
	\begin{equation}
		\lga(s) = C_{\alpha} s^{-\alpha} 
		\exp\left(
			-\alpha\int\limits_s^{\infty} du \frac{\lrho(u)}{u}
		\right),
		\label{eq:g-solution}
	\end{equation}
	where $C_{\alpha}$ is a constant yet to be determined. 

	We now show how to recover $\lSurva(s)$ from $\lga(s)$.  
	We employ the identity
	\begin{equation}
		\int\limits_s^{\infty} du (u - s)^{\beta} e^{-s t} = 
		\Gamma(\beta + 1) t^{-1-\beta} e^{-s t},
		\label{eq:rl-int-exp}
	\end{equation}
	which holds for $\beta>-1$\footnote{
	The integral operator in 
	(\ref{eq:rl-int-exp}) and (\ref{eq:rl-int-lt}) 
	corresponds to the Riemann-Liouville 
	fractional integral. 
	Similarly, the integral in (\ref{eq:surv-from-g}) 
	can be expressed by means of 
	the Caputo fractional derivative.
	}. 
	From (\ref{eq:rl-int-exp}) we deduce that
	\begin{equation}
		\LT\{t^{-\beta-1} f(t)\} = 
		\frac{1}{\Gamma(\beta+1)} 
		\int\limits_s^{\infty} du (u - s)^{\beta} \tilde{f}(s),
		\label{eq:rl-int-lt}
	\end{equation}
	We can apply 
	(\ref{eq:rl-int-lt}) 
	in
	(\ref{eq:g-def}) 
	by replacing $\beta$ with $\alpha-1$ and 
	noticing that $\LT\{t f(t)\} = -\partial_s \tilde{f}(s)$
	\footnote{
	We could have also applied 
	(\ref{eq:rl-int-lt}) 
	directly with $\beta = \alpha - 2$. 
	The resulting equation is simpler, 
	but holds only for $\alpha>1$. 
	}, 
	which leads to
	\begin{equation}
		\lSurva(s) 
		= 
		-\frac{1}{\Gamma(\alpha)} 
		\int\limits_s^{\infty} du 
		(u - s)^{\alpha-1} \partial_u \lga(u).
		\label{eq:surv-from-g}
	\end{equation}
	In the last step we plug (\ref{eq:g-solution}) 
	into (\ref{eq:surv-from-g}). 
	The initial condition 
	\begin{equation}
		\lim_{t\to 0} \Surva(t) = \lim_{s\to\infty}s\lSurva(s) = 1
	\end{equation}
	allows us to calculate $C_{\alpha} = \Gamma(\alpha)$. 
	The final result reads
	\begin{equation}
		\lSurva(s) 
		=
		\alpha
		\int\limits_s^{\infty} du 
		\left(\frac{u - s}{u}\right)^{\alpha-1} 
		\frac{1 - \lrho(u)}{u^2}
		\exp\left(
			-\alpha\int\limits_u^{\infty} \mathrm{d}v \frac{\lrho(v)}{v}
		\right)
		.
		\label{eq:surv-main}
	\end{equation}
	Eq.~(\ref{eq:surv-main}) allows us to calculate any statistics of 
	the completion times, at least in principle. 
	In particular, the \lk{MCT} is simply given by $\lSurva(0)$, i.e.
	\begin{equation}
		\Tm_{\alpha} = 
		\alpha 
		\int\limits_0^{\infty} \dd u
		\frac{1 - \lrho(u)}{u^2}
		\exp\left(
			-\alpha\int\limits_u^{\infty} \dd v \frac{\lrho(v)}{v}
		\right)
		.
		\label{eq:mfht-main}
	\end{equation}
	\subsection{Special cases}
	Here we present two special cases of the search process $\To$, 
	which illustrates basic features of the scale-free resetting. 
	\subsubsection{1D diffusion with a single trap}
\begin{figure}
		\includegraphics[trim={0 0.1cm 1.1cm 0.6cm},clip,width=0.49\linewidth]{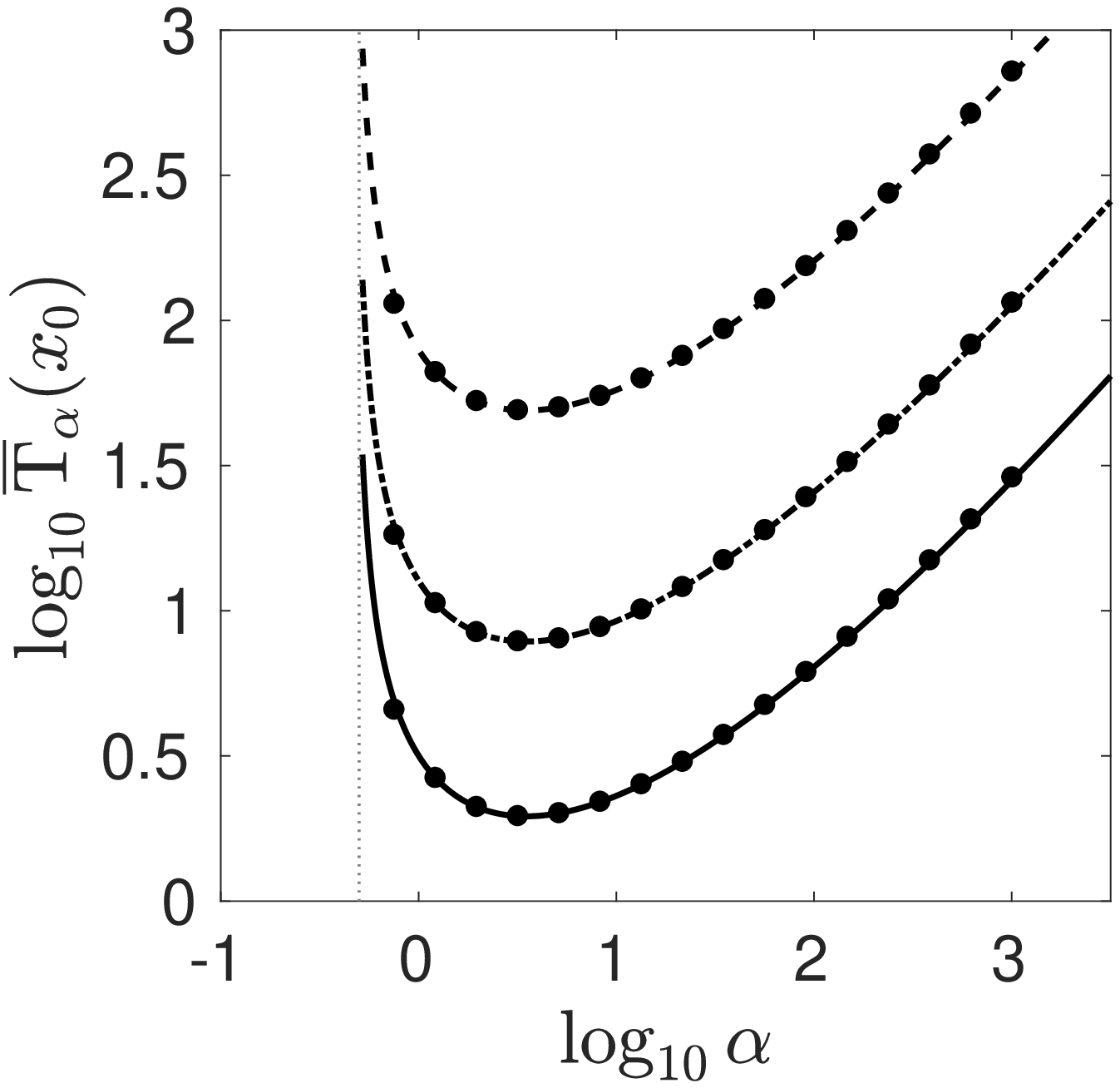}
		\includegraphics[trim={0 0.1cm 1.1cm 0.6cm},clip,width=0.49\linewidth]{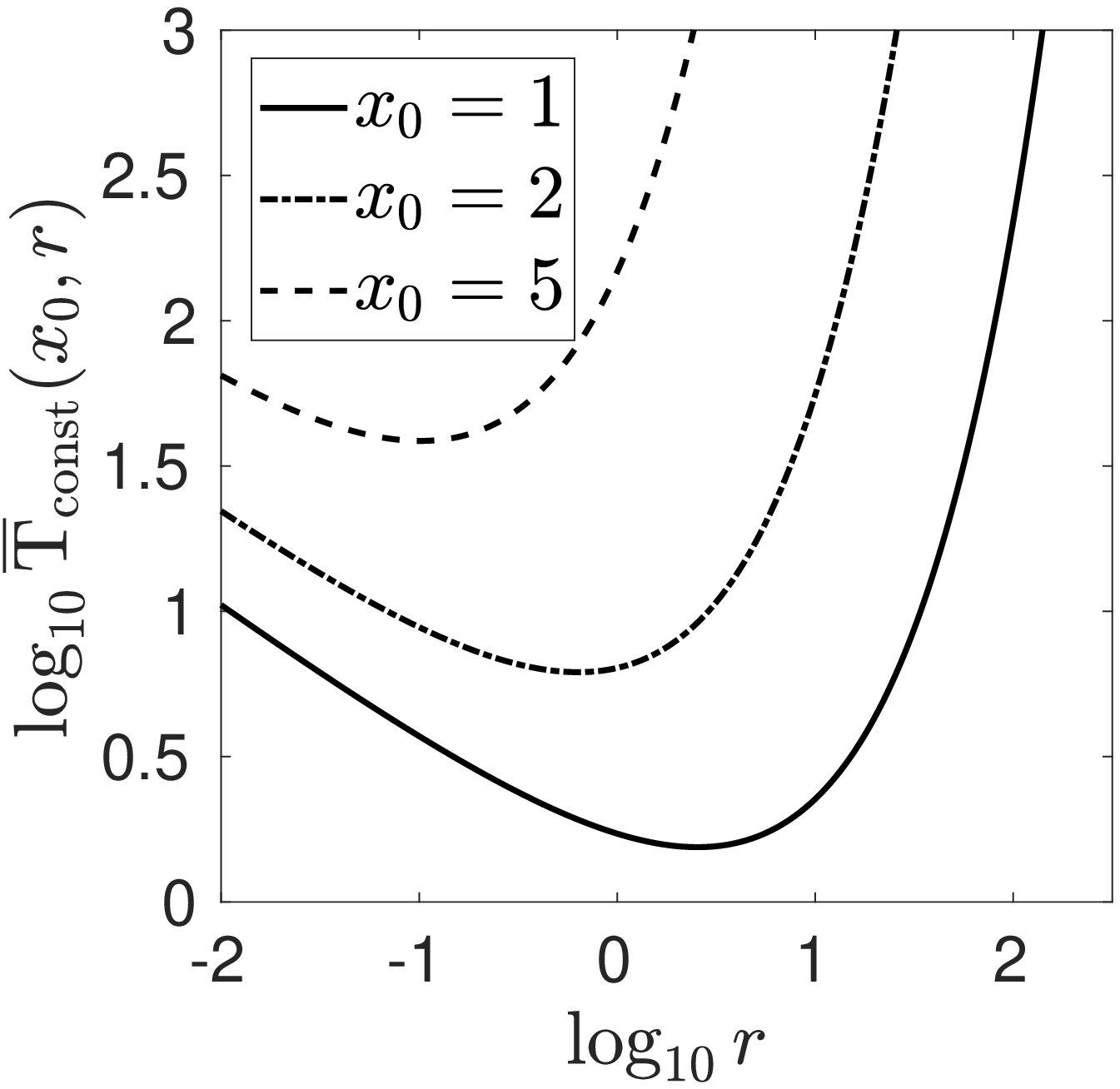}
		\caption{
	    The \lk{MCT} of diffusion with scale-free resetting as a function of $\alpha$ (left) 
		and the \lk{MCT} of diffusion with constant-rate resetting as a function of $r$ (right).
		The gray vertical line denotes the asymptote at $\alpha=\frac{1}{2}$. 
		\lk{Dots denote s}imulation results \lk{which} were obtained by means 
		of the Euler-Maruyama stochastic integration method with a 
		bias reduction technique that modifies the stopping rule close to the target 
		\cite{mannella1999absorbing} with $D=1$, 
		integration time step $\Delta t = 10^{-3}$, 
		and number of samples $M = 10^6$. 
		Notice log-log scale. 
		}
		\label{fig:opt-alpha-diff}
	\end{figure}

	In the first example we consider one-dimensional diffusion 
	(\ref{eq:general-diff-free}) with a single target (trap). 
	The probability density function of the completion times 
	(here corresponding to the first passage times) 
	for $\alpha=0$ (no resets) is well-known and in the Laplace space reads
	\lk{\cite{redner2001guide}}
	\begin{equation}
		\lrho(s) = e^{-\sqrt{s\tdiff}},
		\label{eq:pdf-fpt-free-diff}
	\end{equation}
	where $\tdiff \equiv x^2_0/D$. 
	Importantly, $\rho_0(t)\sim t^{-3/2}$ \lk{for large $t$} 
	and thus the MCT is infinite. 
	Plugging (\ref{eq:pdf-fpt-free-diff}) into the general 
	expression for the \lk{MCT} (\ref{eq:mfht-main}) leads to a 
	rather complicated integral, 
	which can nevertheless be easily integrated numerically. 
	Numerical simulations confirm our theoretical prediction 
	(cf. Fig.~\ref{fig:opt-alpha-diff}) 
	and show that the \lk{MCT} attains its minimum value of 
    $\Tm_{\alpha^*}/\tdiff \approx 1.97$ at $\alpha^*\approx 3.5$. 
    Moreover, as expected from the dimensionality analysis, 
	the \lk{MCT} scales quadratically with the initial distance to the target.
	
	We can learn a lot about the distribution of $\Ta$ by directly 
	analyzing the auxiliary function (\ref{eq:g-solution}), 
	which in the case of diffusion takes the form
	\begin{equation}
		\lga(s) = \Gamma(\alpha+1) s^{-\alpha} 
		\exp\left(
			-2 \alpha E_1\left(\sqrt{s\tdiff}\right)
			\right),
		\label{eq:g-solution-diff}
	\end{equation}
	where
	\begin{equation}
		E_1(x) = 
		\int\limits_x^{\infty}\dd t \frac{e^{-t}}{t}
		\label{eq:exponential-integral}
	\end{equation}
	is a version of the exponential integral. 
	For $x>0$ it can be expressed as
	\begin{equation}
		E_1(x)=  - \gamma - \ln{x} - \sum_{k=1}^{\infty}\frac{(-x)^{k}}{k\cdot k!}
	    \label{eq:E1-expansion}
    \end{equation}
    where $\gamma \approx 0.5772$ stands for the Euler-Mascheroni constant.
	Thus the solution (\ref{eq:g-solution-diff}) has the following representation
	\begin{equation}
		\lga(s) = \Gamma(\alpha+1) \tdiff^{\alpha} 
		\exp\left(
			2 \alpha\gamma + 2\alpha \sum_{k=1}^{\infty}\frac{(-\sqrt{s \tdiff})^{k}}{k\cdot k!}
			\right),
		\label{eq:g-solution-diff-series}
	\end{equation}
	By letting $s=0$ in 
	(\ref{eq:g-solution-diff-series}) 
	we arrive at a simple 
	formula for the $\alpha$-th (fractional) moment 
	\begin{equation}
		\langle \Ta^{\alpha}\rangle = 
		\alpha\lim_{s\to 0}\lga(s) = 
		\Gamma(\alpha+1) e^{2\alpha\gamma} \tdiff^{\alpha}. 
	\end{equation}
	For $\alpha=1$ we obtain \lk{a simple} expression for the MCT
    \begin{equation}
        \Tm_1(x_0) = 
        e^{2\gamma} \tdiff
        \approx 3.17 x_0^2/D
        .
        \label{eq:T-for-alpha-1}
    \end{equation}
	As expected, the \lk{MCT} scales quadratically with the distance to the target.
	Small $s$ expansion of 
	(\ref{eq:g-solution-diff-series}) 
	shows that ${\ga(t)\sim t^{-3/2}}$ for large $t$, 
	which together with (\ref{eq:g-def}) 
	implies that the tail of the survival probability behaves like
	\begin{equation}
		\Surva(t) \sim t^{-1/2 - \alpha}. 
	\end{equation}
	Therefore, the fractional moments ${\langle \Ta^{\nu} \rangle}$ 
	are finite if and only if ${\nu<\alpha + 1/2}$. 
	In particular, the \lk{MCT} is finite for $\alpha>1/2$, 
    see Fig.~\ref{fig:opt-alpha-diff}. 
	
	\subsubsection{Random search with failure}
	\label{section:failure}
	\lk{As we show in Section~\ref{section:linear}, 
	the optimal choice of $\alpha$ does not depend on the search-problem time scale.}
	However, this choice may be sensitive to other features of $\rho_0$. 
	In the second example we explore the resilience of the 
	scale-free resetting to search failures. 
	Let the search problem be described by the density
	\begin{equation}
		\rho_0(t) = p\delta(t-\tau_0),
		\label{eq:density-p}
	\end{equation}
	where $p \leq 1$ is the probability of a successful search. 
	The trivial case $p=1$ corresponds to deterministic completion time, 
	whereas for $p<1$ the density is unnormalized 
	as there is non-zero probability $1-p$ that the search ends with a failure, 
	i.e. the completion time is infinite. 
	We plug  
	\begin{equation}
		\lrho(s) = p e^{-s\tau_0}
	\end{equation}
	into (\ref{eq:g-solution}) and arrive at 
	\begin{equation}
		\lga(s) = \Gamma(\alpha+1) s^{-\alpha} 
		\exp\left(
			-\alpha p E_1\left(s\tau_0\right)
			\right),
		\label{eq:g-solution-p}
	\end{equation}
	and thus via the expansion (\ref{eq:E1-expansion})
	\begin{equation}
		\Surva(t) \sim t^{-\alpha p}
	\end{equation}
    for \lk{$p<1$ and} large $t$.
	We conclude that the \lk{MCT} is finite if and only if 
    \lk{$p=1$ or} $\alpha>1/p$, see Fig.~\ref{fig:opt-alpha-fail}. 
	Moreover, the \lk{MCT} is again proportional to time scale $\tau_0$:
	\begin{equation}
		\Tma(p) = 
		\tau_0 
		\alpha \int\limits_0^{\infty} \dd s
		\frac{1 - p e^{-s}}{s^2} e^{-\alpha p E_1(s)}.
		\label{eq:Ta-fail}
	\end{equation}
	This example illustrates the fact that, 
	while robust to the time scale,
	scale-free resetting may be sensitive to other features of the search problem. 
	In particular, the higher the probability of the search failure,  
	the larger resetting intensity $\alpha$ is necessary to 
	assure a finite value of the MCT. 
	We discuss this issue in more detail in Section~\ref{sec:tradeoff}.
	\subsection{Linear scale-dependence}
    \label{section:linear}
	We have seen that in two simple special cases the \lk{MCT} 
	is proportional to the time scale of the underlying search problem.
	Here we show that this is always the case and that this observation generalizes to 
	higher moments. 
	Given the completion time of a process without resets $\To$, 
	we define a rescaled completion time as $\scale \To$. 
	As in Section~\ref{section:failure}, 
	the parameter $\scale$ denotes the time-scale of the search problem, 
	and may be related to its different features, e.g. 
	in the case of diffusion $\scale$ is proportional to $x_0^2/D$. 
    The corresponding probability density functions are related as
    \begin{equation}
		\lrho(s,\scale) = \langle \exp(-s \scale \To)\rangle_{\To} = \lrho(\scale s,1),
	\end{equation}
	whereas the survival probabilities 
	\begin{equation}
		\lSurv_0(s,\scale) = \scale \lSurv_0(\scale s,1).
	\end{equation}
    Let $\Ta(\scale)$ denote the completion time of a process with 
    scale-free resetting 
	\lk{with the underlying reset-free completion times given by} $\scale \To$. 
    We substitute variables ${v \to v/\scale}$ and ${u \to u/\scale}$ 
	in (\ref{eq:surv-main}) 
    and obtain $\lSurva(s,\scale) = \scale \lSurva(\scale s,1)$, 
	which leads to the conclusion that
    \begin{equation}
        \Ta(\scale) \sim
        \scale \Ta(1),
    \end{equation}
    where $\sim$ denotes the equality of distributions. 
	For all finite fractional moments 
	\begin{equation}
	\langle \Ta(\scale)^{\nu} \rangle \propto \scale^{\nu},
	\label{eq:moments-scaling}
	\end{equation}
	and thus the mean value scales linearly $\Tma(\scale) \propto \scale$:  
    scale-free restarts of the form (\ref{eq:r-propto-inv-t})
    yield \lk{the MCT} that is proportional to 
	time scale of the underlying search problem, 
    which explains why in the case of diffusion $\Tma \propto x_0^2$.
	As a consequence of (\ref{eq:moments-scaling}), the optimal value of $\alpha$, 
	as defined by any moment, is robust to changes of $\scale$.
	\lk{To see this, we can rewrite (\ref{eq:moments-scaling}) in 
	the form 
	$\langle \Ta(\scale)^{\nu} \rangle = \scale^{\nu} h(\alpha)$, 
	where the function $h(\alpha)$ does not depend on $\scale$. 
	Clearly, the value of $\alpha$ that minimizes $h(\alpha)$ 
	at the same time minimizes the $\langle \Ta(\scale)^{\nu} \rangle$ 
	for any non-zero value of $\scale$.
	}
    This result is quite remarkable: 
	the scale-free resetting provides a parsimonious search boosting technique 
    that does not employ any knowledge about the 
	time scale of the underlying search problem. 

	\subsection{Comparison with other resetting protocols}
	\label{sec:comparison}
	\subsubsection{Resetting protocols}
	In order to understand merits and limitations of the proposed 
	scale-free resetting, it is instructive to compare it to other 
	\textit{resetting protocols}, i.e. predefined schemes 
	of introducing resets into the system. 
	We focus solely on (possibly non-stationary) 
	feedforward protocols that are independent 
	from the state of the system
	\footnote{
	In general resetting protocols may form a feedback loop that 
	depend on the state of the system 
	(e.g. current position $x$ in the case of diffusion 
	\cite{pinsky2018diffusive}). 
	}

	Due to its simplicity, constant-rate resetting (i.e. $r(t)=r$) 
	is particularly popular in the literature. 
    Diffusion with such resetting has been studied extensively 
	and its optimal \lk{MCT} is given by 
    $\Tm_{\mathrm{const}}(x_0,r^*) \approx 1.54 x_0^2/D$ 
    with \lk{the optimal resetting rate} 
	$r^*\propto x_0^{-2}$ \cite{evans2013optimal}. 
	
	Another protocol that is of interest to us is a periodic, deterministic resetting.  
	Following \cite{Reuveni2016,pal2016first} we call this resetting \textit{sharp}. 
	Sharp resetting is important, as it was shown to form the dominant strategy 
	in the stationary setting \cite{pal2016first}.

	\subsubsection{Fluctuating or unknown environment} 
	Although the constant resetting can work better than the scale-free resetting 
	($\Tm_{\mathrm{const}}(x_0,r^*) < \Tm_{\alpha^*}(x_0)$), 
	the optimal scale-free search parameter $\alpha^*$ 
	does not involve any knowledge about the distance to the target. 
	In contrast, the optimal constant rate $r^*$ depends strongly on $x_0$, 
    see Fig.~\ref{fig:opt-alpha-diff}. 
	Hence, optimizing $r$ may be hard, or even impossible, 
	if the position of the target is unknown. 
	
	Consider a scenario of a single, immobile target placed at a random position
	and assume the location of the target does not change between the resets, 
	but is drawn independently for separate trials 
	\cite{kusmierz2015optimal, kusmierz2017optimal, pinsky2018optimizing}. 
    This is equivalent to the resetting (and initial) position 
	being drawn from a distribution $\rho_X(x_0)$, 
	i.e., the \lk{MCTs of the scale-free and constant resetting 
	processes in this case can be calculated as 
	$\langle \Tma(x_0) \rangle_{x_0 \sim\rho_X}$
	and
	$\langle \Tm_{\mathrm{const}}(x_0) \rangle_{x_0 \sim\rho_X}$,
	respectively. 
	The d}istribution $\rho_X$ may represent the real variability 
    of the environment or the ignorance of the searching agent. 
	
	As an important special case, let us take the Laplace distribution
	\lk{
	$\rho_X(x_0) = \exp\left(-|x_0|/\lambda\right)/(2\lambda)$, 
	}
	which maximizes entropy for a given average distance to the target 
	$\lambda = \langle |x_0|\rangle$. 
	In this case, the \lk{MCT} of diffusion with constant resetting rate $r$ 
	reads
	\lk{
	\begin{equation}
	\langle \Tm_{\mathrm{const}}(x_0,r) \rangle_{x_0\sim\rho_X} = 
	\frac{\lambda}{\sqrt{D r} - r\lambda}
	\label{eq:const-resets-lambda}
	\end{equation}
	with the minimum value of $4\lambda^2/D$ at $1/r^* = 4\lambda^2/D$. 
	In contrast, 
	since} the variance of the Laplace distribution is equal to $2\lambda^2$, 
	the \lk{MCT} of diffusion with scale-free resetting is given by
	\begin{equation}
		 \langle \Tm_{\alpha}(x_0) \rangle_{x_0\sim\rho_X} = 
		 f(\alpha) \frac{2\lambda^2}{D},
	\end{equation}
	where $f(\alpha)=\Tm_{\alpha}(x_0)/\tdiff$ is the same 
	prefactor as in the case of constant $x_0$ 
	(Fig.~\ref{fig:opt-alpha-diff}) with the minimum 
	${f(\alpha^*)\approx 1.97}$ 
	at ${\alpha^*\approx3.5}$---in this case 
	the scale-free resetting leads to a (slightly) better 
	efficiency than the constant resetting. 
	More importantly, in that case the optimal choice of 
	$\alpha$ does not depend on $\lambda$, 
	to which the optimal constant rate is quite sensitive.
	Indeed, if the chosen constant rate is larger than 
	\lk{$r=D/\lambda^{2}$}, 
	the \lk{MCT} diverges, see (\ref{eq:const-resets-lambda}).  
	
	More broadly, the \lk{MCT} of constant resetting search diverges for 
	any distribution of $x_0$ that has tails heavier than exponential, 
	in particular for any power-law distribution.
	In contrast, the \lk{MCT} of diffusion with scale-free resetting remains finite 
	for any distribution of $x_0$ with finite variance 
	and the choice of the optimal parameter $\alpha$ does not depend on any 
	aspect of the distribution of $x_0$, 
	
	\subsubsection{Trade-off}
	\label{sec:tradeoff}
    Our results suggest that resetting protocols entail a natural trade-off 
	between sensitivity of the protocol outcomes to two distinct features 
	of the completion times of the underlying search problem: 
	its time scale and the shape of its tail. 
	It is convenient, in this context, to analyze the 
	survival probability of the completion times.  
	The tail of the survival probability captures both 
	probability of large fluctuations and the failure probability, 
	i.e. in the case of a non-zero failure probability, 
	the survival probability does not decay to zero. 

\begin{figure}
		\includegraphics[trim={0 0.2cm 1.1cm 0.6cm},clip,width=0.49\linewidth]{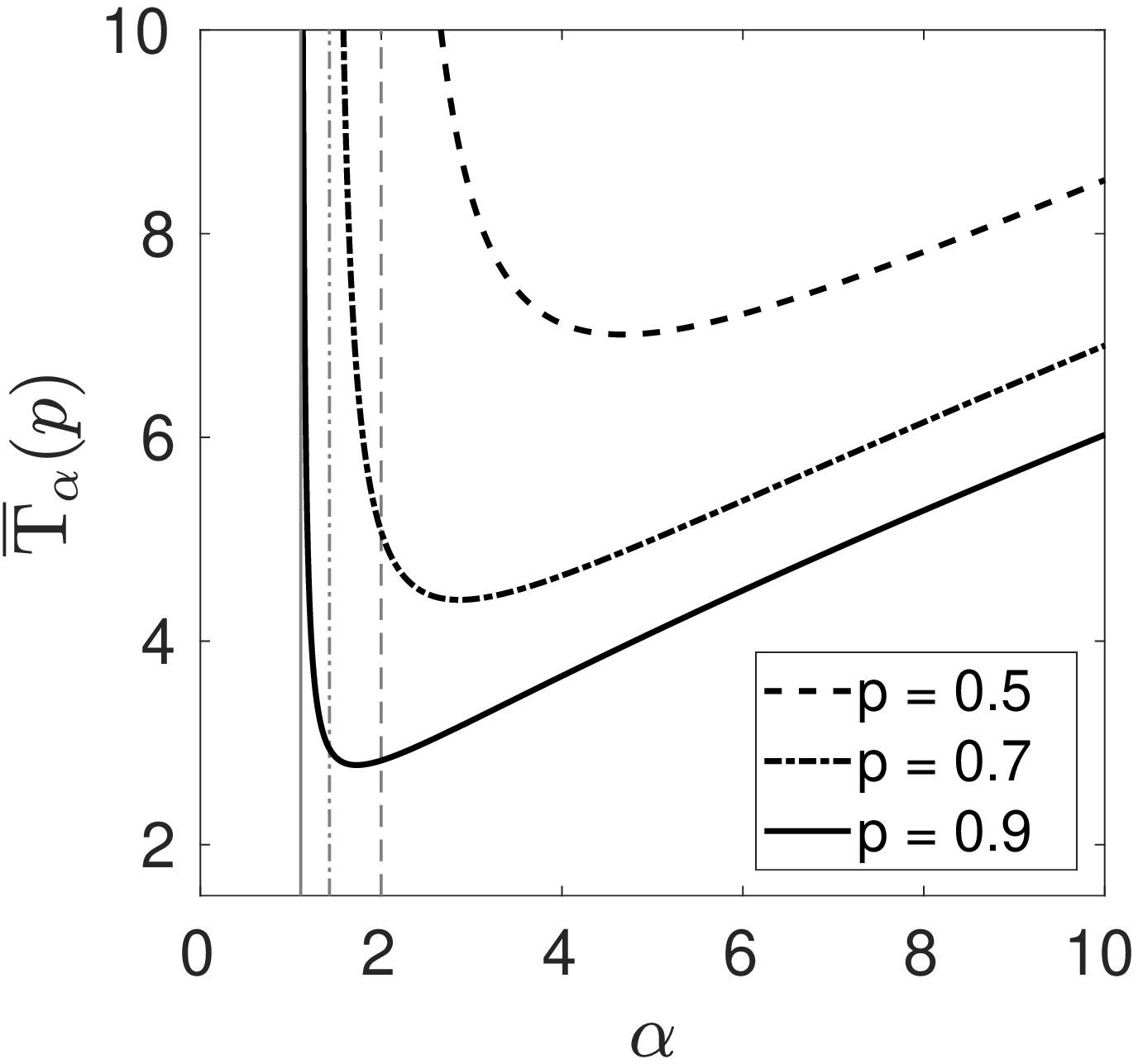}
		\includegraphics[trim={0 0.2cm 1.1cm 0.6cm},clip,width=0.49\linewidth]{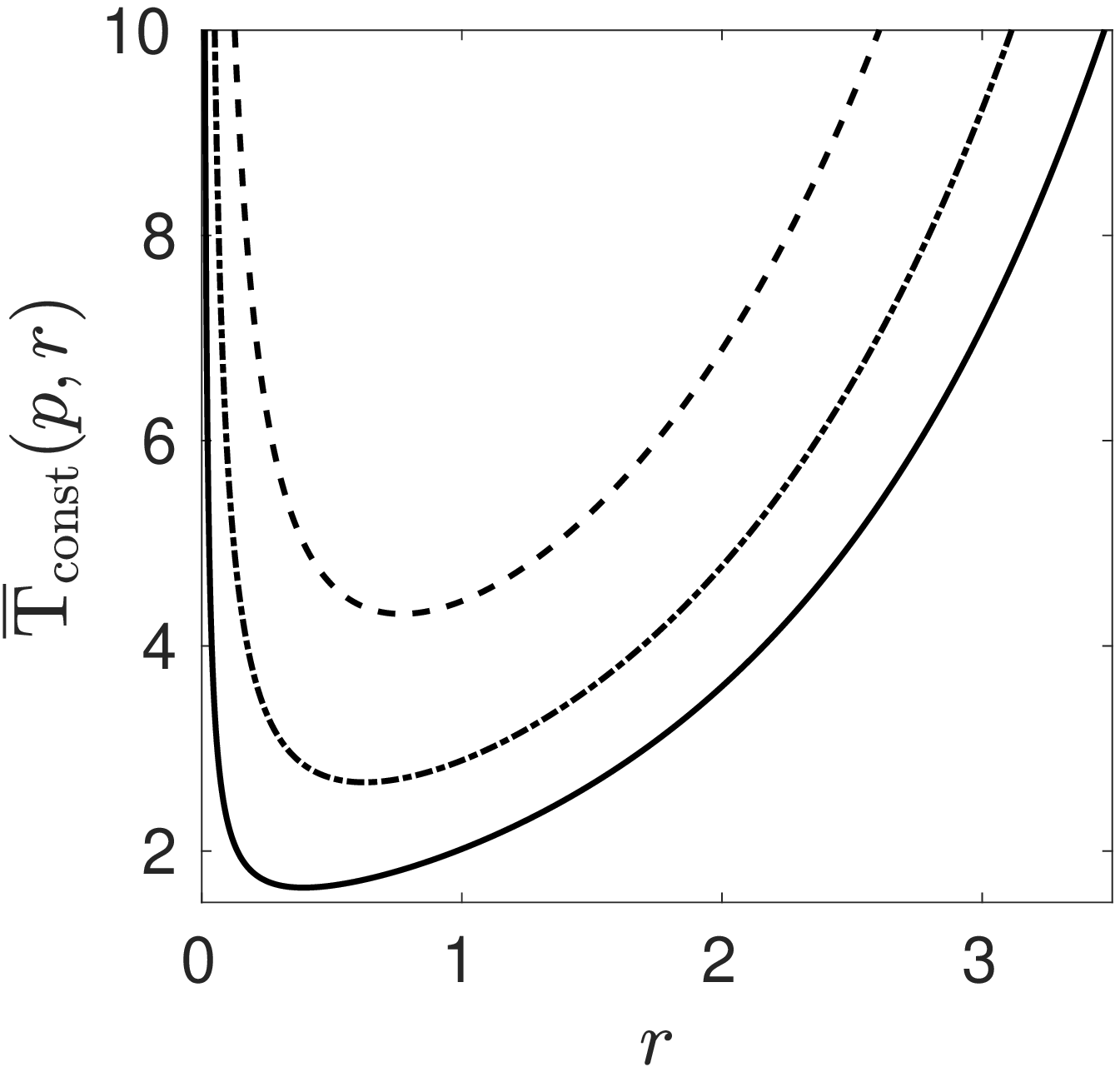}
		\caption{
	    The \lk{MCT} of the failure model (\ref{eq:density-p}) with scale-free resetting 
		as a function of $\alpha$, Eq.~(\ref{eq:Ta-fail}) (left) and
		with constant-rate resetting as a function of $r$, Eq.~(\ref{eq:Tconst-fail}) (right).
		Gray vertical lines denote asymptotes given by $\alpha_{\infty}=1/p$. 
		}
		\label{fig:opt-alpha-fail}
	\end{figure}

	The scale-free resetting is robust to the time scale. 
    However, the time-scale robustness does not come without a price: 
	due to the ever-growing intervals between successing resets, 
	this protocol is rather sensitive to the tail of the distribution. 
	In the case of a one-dimensional diffusive search, 
	the survival probability of the completion times 
	has a power-law tail $\sim t^{-1/2}$. 
	The scale-free resetting to some extent tempers the tail, 
	leading to a finite value of the MCT for $\alpha > \frac{1}{2}$. 
	However, the fluctuations of the completion times are still large, 
	with the survival probability $\sim t^{-1/2-\alpha}$.  
	Thus, the variance diverges for $\alpha\leq\frac{3}{2}$. 
	Similarly, the scale-free resetting is sensitive to the failure probability: 
	as shown above, in order to retain a finite MCT, 
	$\alpha$ has to be larger than $1/p$, 
	where $1-p$ is the failure probability, 
	see Fig.~\ref{fig:opt-alpha-fail}. 

	At the other extreme is the sharp resetting protocol. 
	Indeed, the optimal sharp resetting 
    shows shorter \lk{MCT}s and lower relative fluctuations than any 
	stochastic resetting in a known search environment \cite{pal2016first}. 
    Additionally, it is easy to show that the optimal sharp protocol 
	does not depend on the failure probability $p$.
	However, sharp resets are at the same time extremely sensitive 
	to the time scale of the search process. 
    In the case of diffusion they lead to divergent 
	\lk{MCT}s for any Laplace distribution of $x_0$ \cite{kusmierz2017optimal}.

	Within the framework of the discussed trade-off, constant-rate 
	resetting is placed in between sharp resetting and scale-free resetting. 
	In the case of diffusion, fluctuations of the completion times are 
	strongly reduced by this kind of resetting---all moments are finite for any $r$ 
	\cite{Reuveni2016,pal2016first}. 
	In the simple case of constant time failure model (\ref{eq:density-p}) 
	its \lk{MCT} reads
	\begin{equation}
		\Tm_{\mathrm{const}} = 
		\frac{1}{r}\left( \frac{e^{r \tau_0}}{p} - 1 \right).
		\label{eq:Tconst-fail}
	\end{equation}
	The optimal parameters of the constant-rate resetting protocol depend 
	both on scale and failure probability, see Fig.~\ref{fig:opt-alpha-fail}. 
	On the one hand, in contrast to the scale-free resetting protocol, 
	the \lk{MCT} is in this case finite for any $p$. 
	On the other hand, for a given \lk{value of the} resetting rate, 
	the completion time is very sensitive to $\tau_0$. 

	To sum up, sharp resetting is the most efficient 
	resetting protocol in a well-known, static environment (search problem). 
	However, such a protocol is highly sensitive to the time scale of 
	the search problem and may easily fail in the case of
	an imperfect knowledge or fluctuating environment. 
	If these fluctuations (or ignorance) are not too large and 
	the average time scale is known, constant-rate stochastic resets are advantageous. 
	Otherwise, scale-free resets may be of great advantage, 
	since they are robust to the time scale of the search problem. 
	
	\subsubsection{Generalizations}
	The constant-rate and scale-free resetting protocols are special cases of 
	the general family of resetting protocols of the form $r(t)=\alpha/t^{\mu}$, 
	introduced in Section~\ref{sec:generalizations1}.  
	As shown therein, supplementing diffusion with 
	such a resetting protocol with $\mu<1$ gives rise to the subdiffusive behavior. 
	Thus, this protocol introduces a time scale and, for a fixed $\alpha$, 
	its MCT scales superlinearly with $\scale$.
    It is therefore not robust, as the optimal $\alpha$ depends on $\scale$.
	Nonetheless, this protocol may still be efficient in some search problems.
	Since in the limit of $\mu\to 0$ the standard constant-resetting case is recovered, 
	in search problems with large variability one can expect that the fluctuations 
	of the completion times are smaller for lower values of $\mu$. 
	Thus, the subdiffusive protocol may prove useful in the context of 
	balancing the trade-off discussed in Section \ref{sec:tradeoff}.
	The rich family of models with absolute-time-dependent stochastic 
    resetting and the associated trade-off will be the subject of a separate study.

	\section{Conclusions}
	We envision multiple applications of scale-free resetting, 
	especially in optimization problems.
    For instance, it seems natural to ask about its relation to simulated annealing 
	\cite{kirkpatrick1983optimization}, and its applicability in evolutionary 
	algorithms \cite{fukunaga1998restart,jansen2002analysis},  
	deep learning via gradient methods \cite{lecun2015deep, luersen2004globalized}, 
	and biologically plausible learning techniques with three factors \cite{fremaux2016neuromodulated,kusmierz2017learning}. 
	The robustness of the scale-free resetting protocol should 
	offer additional benefits in the non-stationary setup of curriculum learning 
	\cite{bengio2009curriculum} and could potentially offer an explanation as to why aging 
	\cite{villain1977spin, metzler2014anomalous} may be useful in learning.
	Of course the practical optimization problems are high-dimensional 
    and thus the general renewal framework (\ref{eq:renewal-general}) 
	should prove useful in the construction and analysis of the practical algorithms.
    Another interesting avenue of possible applications are models of 
	evidence accumulation and decision making 
	\cite{bogacz2006physics,tajima2016optimal}, 
	as recently it was shown that stationary resets can modify splitting probabilities 
	\cite{belan2018restart}. 
	Restarts in this context may be interpreted as useful forgetting
	\cite{wixted2004psychology,hardt2013decay}. 

	A number of open problems are left for future studies. 
	In order to assess the efficiency of the proposed scheme, 
	one could compare it to diffusion in time-dependent, scale-free potentials---such 
    comparison in the case of static potentials and resets seems to 
	favor the latter \cite{evans2013optimal,kusmierz2017optimal}. 
	Moreover, it is important to find conditions under which scale-free resets 
	can lower the expected completion time, similar to the simple criterion 
	in the case of constant-rate resets 
	\cite{reuveni2014role,rotbart2015michaelis,Reuveni2016}. 
	A related question is how the optimal $\alpha$ 
	depends on the details of a search problem at hand. 
	Another important issue is the divergence of (\ref{eq:r-propto-inv-t}) at $t=0$ 
	which is infeasible and in practice a short-time cut-off has to be introduced. 
	The optimal cut-off should depend on a cost associated with resets.  

	\begin{acknowledgments}
	\lk{We are indebted to Shun Ogawa for stimulating discussions. 
	This work was supported by RIKEN Center for Brain Science, 
	Brain/MINDS from AMED under Grant Number JP19dm020700 
	and JSPS KAKENHI Grant Number JP18H05432.
	}
	\end{acknowledgments}

    \bibliography{citations}
    \bibliographystyle{phjcp}

\end{document}